\documentclass[12pt]{article}
\usepackage{amsmath,amssymb,epsfig}
\pdfoutput=1

\usepackage{color}
\input{colordvi.tex}
\def\unit{{\relax{\rm 1\kern-.26em I}}}

\makeatletter

\renewcommand\section{\@startsection {section}{1}{\z@}%
                                   {-3.5ex \@plus -1ex \@minus -.2ex}%
                                   {2.3ex \@plus.2ex}%
                                   {\normalfont\large\bfseries}}

\renewcommand\subsection{\@startsection{subsection}{2}{\z@}%
                                     {-3.25ex\@plus -1ex \@minus -.2ex}%
                                     {1.5ex \@plus .2ex}%
                                     {\normalfont\normalsize\bfseries}}

\makeatother

\begin{document}

\baselineskip=18pt  
\numberwithin{equation}{section}  
\allowdisplaybreaks  



%
%


\thispagestyle{empty}

\vspace*{-2cm}
\begin{flushright}
\end{flushright}

\begin{flushright}

\end{flushright}

\begin{center}

\vspace{1.4cm}

\vspace{1cm}
{\bf\Large Light Higgsino as the tail of the $\mu$-$B_\mu$ solution}

\vspace*{1.3cm}

{\bf
Masaki Asano$^{1}$ and Norimi Yokozaki$^{2}$} \\
\vspace*{0.5cm}

${ }^{1}${\it Physikalisches Institut and Bethe Center for Theoretical Physics, Universit\"at Bonn, Nussallee 12, D-53115 Bonn, Germany}\\
${ }^{2}${\it Istituto Nazionale di Fisica Nucleare, Sezione di Roma, Piazzale Aldo Moro 2, I-00185 Rome, Italy}\\

\vspace*{0.5cm}

\end{center}

\vspace{1cm} \centerline{\bf Abstract} \vspace*{0.5cm}

Gauge mediation predicts 10\,TeV or heavier squarks because such a heavy stop is required to explain the observed 125 GeV Higgs boson mass without a large trilinear soft mass term in the minimal supersymmetric standard model. Although such a high scale cannot be searched by the LHC directly, gauge mediation also predicts a hierarchy $\mu^2 \ll B_\mu$ by 
a simple and naive solution of the $\mu$ problem. 
We point out that this simple and naive way of generating the $\mu$ ($B_\mu$) term works
 in the case of 10\,TeV or heavier squarks with a slight breaking of a GUT relation among messenger $B$-terms (or supersymmetric mass terms). 
Furthermore, the upper bound on the Higgsino mass is obtained from the observed Higgs boson mass, perturbativity of a relevant coupling, and  conditions avoiding tachyonic sneutrinos and stop. It turns out that the light Higgsino of $\mathcal{O}(100)$\,GeV is a promising signal of gauge mediation.

\newpage
\setcounter{page}{1} 



\section{Introduction}

LHC experiments search the physics beyond the standard model (SM) and one of the promising candidates is supersymmetry (SUSY), which can solve the hierarchy problem. Theory with gauge mediated SUSY breaking~\cite{gmsb,gmsb_old,gmsb_reviews} is an interesting scenario because it can naturally suppress dangerous flavor changing neutral currents.

One of the generic features of gauge mediation is small trilinear $A$ terms. The stop radiative corrections to the lightest Higgs boson mass~\cite{higgs_rad} is, however, maximized if the $A$ term is as large as the stop masses. If the $A$ term is much smaller than stop masses, relatively large stop masses are required to explain the observed Higgs boson mass. Therefore, in gauge mediation, the observed 125 GeV Higgs boson predicts relatively large squark masses, e.g. $m_{\tilde{q}} \sim 10$ TeV with $\tan\beta \sim 10$ ($m_{\tilde{q}} \sim 10^3$ TeV with $\tan\beta \sim 2$).

This prediction will drastically change the difficulty to solve the $\mu$-$B_\mu$ problem. The $\mu$ problem can be solved by considering a mechanism of the generating $\mu$ term from the SUSY breaking. Such a mechanism, however, usually provides also a very large $B_\mu$ term, $\mu^2 \ll B_\mu$, in gauge mediation. This is called the $\mu$-$B_\mu$ problem~\cite{muBmuref01,muBmuref02}: The $\mu^2 \ll B_\mu \sim m_{\tilde{q}}^2$ spectrum is required in order to achieve the electroweak symmetry breaking (EWSB), while such a very light Higgsino have already excluded if $m_{\tilde{q}} \sim \mathcal{O}(1)$ TeV. 
However, once we suppose such a relatively high SUSY scale, $m_{\tilde{q}} \sim 10$ TeV, even the light Higgsino mass can be $\mathcal{O}(100)$ GeV and has not been excluded. 
Thus, this hierarchical relation, $\mu^2 \ll B_\mu$, may be no longer a problem and will be a prediction of the light Higgsino.\footnote{
In Ref.~\cite{Cohen:2015lyp}, it has been suggested that the $\mu$-$B_\mu$ problem is solved simply in mini-split SUSY spectra with stop mass $\gtrsim 100$ TeV. 
}

In this paper, we investigate the phenomenology expected from such a simple solution of the $\mu$-$B_\mu$ problem. In particular, we have interested in the $m_{\tilde{q}} \sim 10$\,TeV region rather than a region of $10^3$\,TeV or heavier squarks because the naturalness of achieving the correct EWSB is much better than that region.

We point out that such a simple and naive $\mu$-$B_\mu$ solution works considering a slight breaking of a grand unified theory (GUT) relation in the messenger sector.\footnote{
Actually, such spectra, $\mu^2 \ll B_\mu \sim m_{\tilde{q}}^2 \sim (10 ~{\rm TeV})^2$ with $\tan\beta \sim 10$, are not consistent with the EWSB conditions in the minimal messenger model. Remember that the next-to-lightest SUSY particle is not Higgsino but either bino or stau in this case.
}
The doublet/triplet splitting can provide such a small $\mu$ term by a cancellation in Higgs soft masses through radiative corrections. This focusing effect \cite{Agashe:1997kn,Agashe:1999ct,Cheung:2007es} is compatible with the hierarchical spectra of the simple $\mu$-$B_\mu$ solution, $\mu^2 \ll B_\mu \sim m_{\tilde{q}}^2 \sim (10 ~{\rm TeV})^2$ with $\tan\beta \sim 10$.

We also suggest that the size of $\mu$ is bounded from above. Taking into account the perturbativity up to the GUT scale and avoiding tachyonic sneutrinos and stop, we show that the light Higgsino of $\mathcal{O}(100)$\,GeV is a promising signal in gauge mediation.

\section{Generating $\mu$ term and light Higgsino prediction}

To address the $\mu$ problem, we assume that the $\mu$ term is initially forbidden in the superpotential due to a symmetry, then it is generated by SUSY breaking. It can be realized if the Higgs superfields are coupled with the messenger sector in gauge mediation. 

As a simple and concrete model, we consider the following superpotential which includes Higgs-messenger couplings, 
\begin{eqnarray}
W &=& -\xi_Z Z + (M_{\rm mess} + k Z) \Psi \bar \Psi, \nonumber \\
 &+& \lambda_u H_u {\bar \Psi}_L \bar N + \lambda_d {\Psi}_L H_d  { N} + M_N N \bar N , 
\label{model}
\end{eqnarray}
where $\Psi$ ($\bar \Psi$) is the messenger superfield transforming ${\bf 5}$ (${\bf \bar 5}$) in SU(5) GUT gauge group and can be decomposed as $\Psi = \Psi_L + \Psi_D$ ($\bar \Psi = \bar \Psi_L + \bar \Psi_D$). The mass parameters, $M_{\rm mess}$ and $M_{N}$, are $R$-symmetry breaking parameters with $R$-charge 2 and we assume their origins are the same, then $M_{\rm mess} \sim M_N$. The breaking of $R$-symmetry is essential  to generate non-vanishing gaugino masses. 
From interactions in the first and second line of Eq.\eqref{model}, $R$-charges of $H_u$ and $H_d$ are fixed as $Q(H_u) + Q(H_d)=4$, which forbids the bare $\mu$-term. It is assumed that $k \left<Z\right>  \ll M_{\rm mess}$ which is ensured by e.g. $K=-|Z|^4/M_*^2$ in the K{\" a}hler potential ($M_*^2 \ll  M_{\rm mess} M_P$ with $M_P=2.4\cdot10^{18}$\,GeV). For simplicity, we drop $Z N \bar N$ term in the following discussions.

As we will discussed in the next section, a slight violation of the GUT relation of messengers is essential for achieving the correct EWSB. Therefore, we define the messenger sector as 
\begin{eqnarray}
(M_{\rm mess} + k Z) \Psi \bar \Psi \to (M_L + k_L Z) \Psi_L {\bar \Psi}_L + (M_D + k_D Z) \Psi_D {\bar \Psi}_D.
\end{eqnarray}
Then, soft SUSY breaking mass parameters are determined by $\Lambda_L = k_L \left<F_Z\right>/M_L$ and $\Lambda_D = k_D \left<F_Z\right>/M_D$, where $\left<F_Z\right> =  \xi_Z$. 
If the GUT relations, $k_L=k_D$ and $M_L=M_D$ at the GUT scale, are satisfied, $\Lambda_L=\Lambda_D$ at any scale. 
These messengers mediate the SUSY breaking to the minimal supersymmetric standard model (MSSM) sector. The explicit formulas for the soft SUSY breaking masses are shown in Appendix A.

After integrating out messengers, $N$ and $\bar N$, not just $\mu$ term but also $B_\mu$-term and other Higgs soft masses are generated.  
The leading and subleading contributions are given as
\begin{eqnarray}
\mathcal{L} &\ni& 
- m_{H_u}^2 H_u^\dagger H_u - m_{H_d}^2 H_d^\dagger H_d 
\nonumber \\ &&+
\left( 
B_\mu H_u H_d 
- A_{u} H_{u} \frac{\partial }{\partial H_u} W 
- A_{d} H_{d} \frac{\partial }{\partial H_d} W    
+ \int d^2\theta \mu H_u H_d
 \right )+h.c.,
\nonumber
\end{eqnarray}
\begin{eqnarray}
m_{H_{u,d}}^2 &=&  \frac{\lambda_{u,d}^2}{16\pi^2} \Lambda_L^2
\Bigl[  P_1(x) + \frac{\Lambda_L^2}{M_L^2} P_2(x) \Bigr], \nonumber \\
\mu &=& \frac{\lambda_{u} \lambda_d  }{16\pi^2} \Lambda_L
\Bigl[  Q_1(x)  + \frac{\Lambda_L^2}{M_L^2} Q_2(x)\Bigr], \nonumber \\
B_\mu &=&  \frac{\lambda_{u} \lambda_d  }{16\pi^2} \Lambda_L^2 
\Bigl[ R_1(x) + \frac{\Lambda_L^2}{M_L^2} R_2(x)\Bigr], \nonumber \\
A_{u,d} &=&  \frac{\lambda_{u,d}^2 }{16\pi^2} \Lambda_L 
\Bigr[ S_1(x) + \frac{\Lambda_L^2}{M_L^2} S_2(x) \Bigr], \label{eq:softparams}
\end{eqnarray}
where we use the same character to denote the Higgs superfields and the scalar components.
The loop functions of the leading contributions are written as 
\begin{eqnarray}
P_1(x) &=& \frac{x^2}{(x^2-1)^3} \left[2(1-x^2)+(1+x^2) \ln x^2 \right] ,\nonumber \\
Q_1(x) &=& \frac{x}{(x^2-1)^2} \left[(x^2-1) -x^2 \ln x^2 \right] ,\nonumber \\
R_1(x) &=& \frac{-x}{(x^2-1)^3} \left[ 1-x^4  + 2x^2 \ln x^2\right],  \nonumber \\
S_1(x) &=& \frac{-1+x^2-x^2 \ln x^2}{(x^2-1)^2},
\end{eqnarray}
with $x = M_N/M_L$.  In the limit of $x=1, \, P_1(1)=1/6, \, Q_1(1)=-1/2, \, R_1(1)=1/3, \, S_1(1)=-1/2$.
The loop-functions of the subleading terms are
\begin{eqnarray}
P_2(x) &=& \frac{1+9x^2-9x^4-x^6+6(x^2+x^4)\ln x^2}{6(x^2-1)^5} , \nonumber \\
Q_2(x) &=&  \frac{-x(2 +3x^2 -6x^4 +x^6 +6x^2 \ln x^2)}{6(x^2-1)^4}  , \nonumber \\
R_2(x) &=& \frac{x(-3-10 x^2 + 18 x^4-6x^6 + x^8 -12 x^2 \ln x^2)}{6(x^2-1)^5}, \nonumber \\ 
S_2(x) &=& \frac{-2-3x^2+6x^4 -x^6 - 6 x^2 \ln x^2}{6(x^2-1)^4},
\end{eqnarray}
which are numerically smaller than leading ones, and can be safely neglected unless $\Lambda_L/M_L$ is very close to 1. 
Here and hereafter, we take $x=1$ as a reference value.

Neglecting the subleading contributions, generating $\mu$ and Higgs soft masses can be written by
\begin{eqnarray}
\mu &\approx& -\frac{1}{2} \left( \frac{\lambda_{u} \lambda_d  }{16\pi^2} \right) \Lambda_L, \quad
B_\mu \approx  \frac{1}{3} \left( \frac{\lambda_{u} \lambda_d  }{16\pi^2} \right) \Lambda_L^2,
\nonumber \\
m_{H_{u,d}}^2 &\approx&  \frac{1}{6} \left( \frac{\lambda_{u,d}^2}{16\pi^2} \right) \Lambda_L^2, \quad
A_{u,d} \approx -\frac{1}{2} \left( \frac{\lambda_{u,d}^2 }{16\pi^2} \right) \Lambda_L.  
\label{eq:softparams_app}
\end{eqnarray}
The difference between $\mu$ and $B_\mu$ is $B_\mu \approx -(2/3) \mu \Lambda_L$.

These generated soft masses and $\mu$-parameter should be consistent with the vacuum conditions. In MSSM, the tree level vacuum conditions are  
\begin{eqnarray}
\frac{m_Z^2}{2}
&=& - |\mu|^2 + \frac{m_{H_d}^2 - m_{H_u}^2 \tan^2\beta}{\tan^2\beta - 1}, 
\label{eq:vacuumcondition1} 
\\
\sin2\beta &=& \frac{2 B_\mu}{m_{H_u}^2 + m_{H_d}^2 + 2|\mu|^2}, 
\label{eq:vacuumcondition2} 
\end{eqnarray}
where $m_Z$ is the $Z$ boson mass. 
From Eq.~\eqref{eq:vacuumcondition2}, in the large $\tan\beta$ case, we obtain
\begin{eqnarray}
    B_\mu \tan\beta \approx m_{H_d}^2  
\quad \to \quad 
\tan\beta \approx \frac{1}{2}\frac{\lambda_d}{\lambda_u}, 
\label{eq:ewsb_tanb}
\end{eqnarray}
therefore, the value of $\mu$ is determined by 
\begin{eqnarray}
|\mu| \approx \frac{1}{4}\frac{\lambda_d^2}{16\pi^2} \frac{\Lambda_L}{\tan\beta} \approx 158\, {\rm GeV}  
\left(\frac{\tan\beta}{10}\right)^{-1}
\left(\frac{\lambda_d}{1.0}\right)^{2}
\left(\frac{\Lambda_L}{1000\,{\rm TeV}}\right). 
 \label{eq:generated_mu}
\end{eqnarray}

The lightest Higgs boson mass depends on $\Lambda_L$ and $\tan\beta$. The lighter squark masses, the larger $\tan\beta$ is required to obtain the observed 125 GeV Higgs mass. Although the details of the dependence will be shown below, actually, in $m_{\tilde{q}} \sim 10$ TeV region of our interest, relatively large $\tan\beta \sim 10$ is required.
Therefore, once we fix the value of $\Lambda_L$, the value of $\mu$ is determined by the $\lambda_d$ by Eq.~\eqref{eq:generated_mu}.

\begin{figure}[!t]
\begin{center}
\includegraphics[scale=1.2]{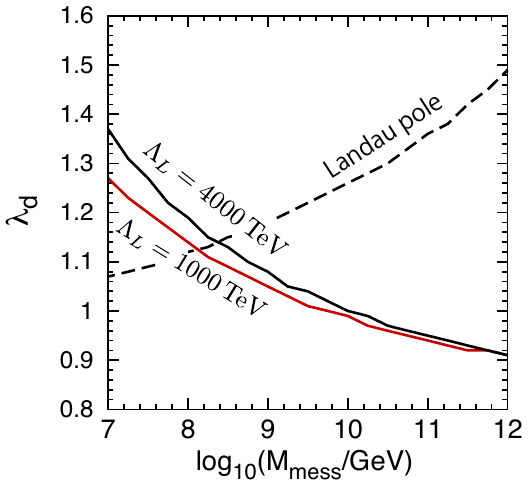}
\caption{
The upper bound on $\lambda_d$ from the Landau pole constraint (dashed-line) and the constraints from the sneutrino mass for $\Lambda_L=1000$ TeV and 4000 TeV with $\Lambda_D=\Lambda_L$ (solid-line).  Above the solid lines, the sneutrinos become tachyonic. We take $\lambda_u = \lambda_d/(2\tan\beta)$. Here, $\alpha_s(M_Z)=0.1185$ and $m_t({\rm pole})=173.34$ GeV.
}
\label{fig:tachyonic_sneutrino}
\end{center}
\begin{center}
\includegraphics[scale=1.2]{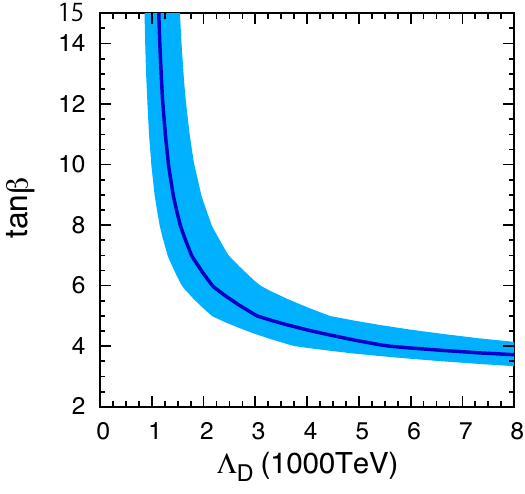}
\caption{
The region consistent with the Higgs mass of 125 GeV, for $\Lambda_L=\Lambda_D$ and $M_{\rm mess}=10^7$ GeV. 
We take $\lambda_u=\lambda_d=0$ and $\mu=-200$ GeV. 
The blue band indicates 
a theoretical uncertainty including the experimental error of the top quark mass ($m_t({\rm pole})=173.34 \pm 0.76$ GeV).
The other parameters are same as in Fig. 1.
%
}
\label{fig:higgs_mass}
\end{center}
\end{figure}

Note that the value of $\lambda_d$ is bounded from above. 
One of the bounds comes from the requirements for avoiding the Landau pole below the GUT scale. We show the upper bound on $\lambda_d$ from the Landau pole constraint by the dashed-line in Fig.~\ref{fig:tachyonic_sneutrino}. The Landau pole constraint is obtained for each $M_{\rm mess}(=M_L=M_D)$ by using one-loop renormalization group equations (RGEs) (shown in Appendix B), demanding that couplings be perturbative up to the GUT scale. 
The other is the constraint from the tachyonic sneutrinos, which caused by non-vanishing $U(1)_Y$ contribution, 
\begin{eqnarray}
&& \frac{d m_L^2}{d \ln \mu_R} \ni \frac{g_1^2}{16\pi^2}\frac{3}{5}(m_{H_d}^2 - m_{H_u}^2) 
\quad \to \quad 
\Delta m_L^2 \simeq -\frac{3}{5} \frac{g_1^2}{16\pi^2} m_{H_d}^2 \ln \frac{M_{\rm mess}}{m_{H_d}} \nonumber \\ && \qquad \qquad \qquad \qquad \qquad \qquad \qquad \qquad \qquad ~~~ 
\approx -(0.11\,\mathchar`-\, 0.36)\lambda_d^2 \frac{\Lambda_L^2}{(16\pi^2)^2}
\label{eq:u1y},
\end{eqnarray}
for $M_{\rm mess}=10^7$-$10^{12}$\,GeV with $\lambda_d \gg \lambda_u$. The above estimation implies $\lambda_d$ can not be much larger than unity, otherwise the negative $U(1)_Y$ contribution becomes larger than the contribution from gauge mediation. We show also the tachyonic sneutrino bounds in Fig.~\ref{fig:tachyonic_sneutrino} using {\tt SOFTSUSY\,3.6.2}~\cite{softsusy} to evaluate MSSM mass spectra. By combining the Landau pole constraint and tachyonic sneutrino constraint, the value of $\lambda_d$ should be
$\lambda_d \lesssim 0.9$-$1.1$ in whole parameter space. 

In Fig.~\ref{fig:higgs_mass}, we show the region consistent with the observed Higgs boson mass on the $\tan\beta$-$\Lambda_D$ plane. The Higgs mass is computed using {\tt SUSYHD\,1.0.2}~\cite{susyhd}.  The blue band indicates theoretical uncertainty, including the experimental error of the top mass, $m_{t}({\rm pole}) = 173.34 \pm 0.76$\,GeV~\cite{topmass}. From the Higgs boson mass constraint, $\tan\beta$ is bounded from below for fixed $\Lambda_D$.

As a result, for the fixed $\Lambda_L$, the upper bound on the Higgsino mass can be obtained from the upper bound on $\lambda_d$ and lower bound on $\tan\beta$ discussed above. From Eqs.(\ref{eq:ewsb_tanb}) and (\ref{eq:generated_mu}), the upper bounds on the Higgsino mass are obtained as
\begin{eqnarray}
|\mu| &\lesssim& (158, 575, 1580)\,{\rm GeV}\,\, {\rm for}\,\, \Lambda_L = (1000, 2000, 4000)\,{\rm TeV},
 \label{mu_bound_0}
\end{eqnarray}
where $\lambda_d=1.0$ and $\Lambda_L=\Lambda_D$ is assumed. 
At the large $\Lambda_L$ case, the value of $\mu$ is pushed up by not only the large $\Lambda_L$ but also small $\tan\beta$, which is required to satisfy $125$ GeV Higgs mass. 
However, as shown in the next section, such a small $\tan\beta$ and large $\lambda_d$ region is constrained requiring 
consistency with the EWSB conditions. As a result, the Higgsino is likely to be light as $\mathcal{O}(100)$\,GeV, at least, in this model.

\section{Electroweak symmetry breaking}
In this section, we point out that a simple and naive way to generate $\mu$-$B_\mu$ term discussed in previous section actually works for the soft masses around 10 TeV, with a slight breaking of a GUT relation among messenger $B$-terms.

At first, we demonstrate the difficulty to satisfy the EWSB condition Eq.~\eqref{eq:vacuumcondition1} by this simple $\mu$ generating mechanism with a simple messenger sector. 
Considering $1/\tan^2\beta \ll 1$ case, the EWSB condition can be written as 
\begin{eqnarray}
\mu^2 \simeq -\frac{m_Z^2}{2} 
- \left( m_{H_u}^2 - \frac{m_{H_d}^2}{\tan^2\beta} \right) + ({\rm CW}), 
 \label{eq:ewsb}
\end{eqnarray}
where (CW) denotes a contribution from Coleman-Weinberg potential.
The Higgs soft masses at the messenger scale are provided by the generating $\mu$ mechanism and usual gauge mediation. Using Eqs.~\eqref{eq:softparams_app} and~\eqref{eq:ewsb_tanb}, it is written by 
\begin{eqnarray}
\left( m_{H_u}^2 - \frac{m_{H_d}^2}{\tan^2\beta} \right)_{\rm mess} \simeq \left[-\frac{\lambda_d^2}{8 \tan^2\beta} 
+ \frac{3}{2}\frac{g_2^4}{16\pi^2} \left(1- \frac{1}{\tan^2\beta}\right) \right] \frac{\Lambda_L^2}{16\pi^2},
\end{eqnarray}
which is negative for large $\lambda_d$ and small $\tan\beta$. 
Additionally, there are radiative corrections from stop and gluino loops, estimated as
\begin{eqnarray}
(\Delta m_{H_u}^2)_{\rm stop/gluino} &\approx& - (0.3\, \mathchar`- \, 0.5) m_{\tilde q}^2 - (0.1\,\mathchar`-\,0.5) M_{\tilde g}^2 \nonumber \\
&\approx& - (0.8\, \mathchar`- \, 1.8)\frac{g_3^4}{(16\pi^2)^2} \Lambda_D^2 ,
\end{eqnarray}
depending on the messenger scale ($M_{\rm mess}=10^7\,\mathchar`-\,10^{12}$\,GeV), 
and also from wino loops and a $U(1)_Y$ contribution. Therefore, Eq.~\eqref{eq:ewsb} can be rewritten as 
\begin{eqnarray}
\mu^2 \simeq -\frac{m_Z^2}{2} 
- \left[ \left( m_{H_u}^2 - \frac{m_{H_d}^2}{\tan^2\beta} \right)_{\rm mess}
+ (\Delta m_{H_u}^2)_{\rm stop/gluino} + (0.1\,\mathchar`- \,0.2) M_{\tilde w}^2  + \Delta_{U(1)_Y}
                    \right] + ({\rm CW}), ~~
 \label{eq:ewsb_corr}
\end{eqnarray}
where $\Delta_{U(1)_Y}$ is the same as Eq.\,(\ref{eq:u1y}) but with an opposite sign. 
The right hand side of Eq.~\eqref{eq:ewsb_corr} is positive and its size is much larger than $\mu^2 \sim (100\,{\rm GeV})^2$ for $\Lambda_L \simeq \Lambda_D$. 
As a result, the EWSB conditions can not be satisfied by this simple $\mu$ generating mechanism with a simple messenger sector. 

\begin{figure}[!t]
\begin{center}
\includegraphics[scale=1.2]{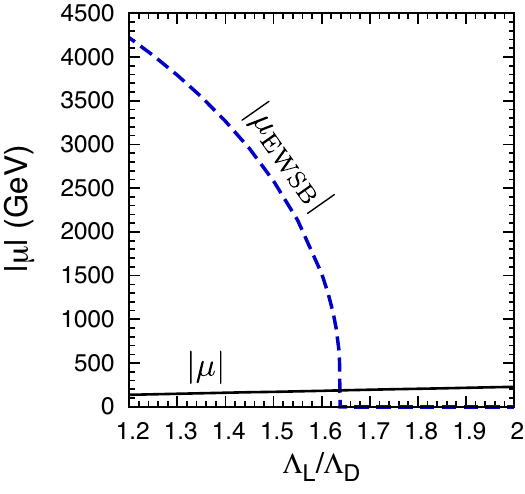}
\caption{
$\mu$ and $\mu_{\rm EWSB}$ as functions of $r_L=\Lambda_L/\Lambda_D$.
We take 
$\Lambda_D=1.5 \cdot 10^6$\,GeV, $M_{\rm mess}=5\cdot 10^6$\,GeV, $\tan\beta=10$, and $\lambda_d=0.7$.
The other parameters are same as in Fig. 1.
%
}
\label{fig:mu}
\end{center}
\end{figure}

The difficulty can be solved by a breaking of a GUT relation with $r_L \equiv \Lambda_L/\Lambda_D > 1$, which enhances the contributions from the wino-loops and $\Delta_{U(1)_Y}$.  
This is similar to a setup of the focus point gauge mediation~\cite{fpgmsb}. Note that $r_L$ is a RGE invariant quantity. 
The correct EWSB can be achieved with $r_L=1.6\,$-$\,2.0$ depending on $\lambda_d$ and $\tan\beta$. 
In Fig.~\ref{fig:mu}, we show the value of the generated $\mu$ term (see Eq.(\ref{eq:softparams})) and the required value $\mu_{\rm EWSB}$ from the EWSB condition, Eq.\,(\ref{eq:ewsb}), as a function of $r_L$. 
The correct EWSB occurs at a point where two-lines of $\mu$ and $\mu_{\rm EWSB}$ cross. It can be seen that the correct EWSB is explained for $r_L \sim 1.6$. 

One might think that by taking large $r_L$, the correct EWSB is always explained for any $\lambda_d$. 
However, this is not true: much larger $r_L$ causes tachyonic stop because radiative corrections give a large negative correction to the right-handed stop mass squared. 
The negative corrections come from a $U(1)_Y$ contribution similar to Eq.~(\ref{eq:u1y}) and Yukawa interactions with top Yukawa coupling because of larger $m_{Q_3}^2$ due to large $SU(2)_L$ contributions. 

\begin{figure}[!t]
\begin{center}
\includegraphics[scale=1.2]{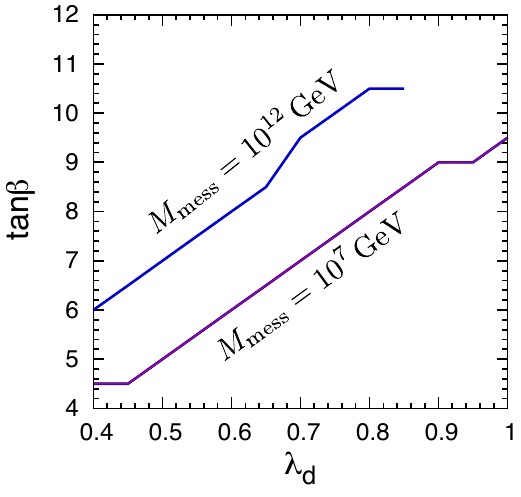}
\caption{
The lower bounds on $\tan\beta$ for $M_{\rm mess}=10^7$\,GeV and $10^{12}$\,GeV from the stop mass.
Below the line, the right-handed stop becomes tachyonic.
We take $\Lambda_D=1000$\,TeV.
The other parameters are same as in Fig. 1.
%
}
\label{fig:neg_stop}
\end{center}
\end{figure}

In Fig.~\ref{fig:neg_stop}, we show the lower bound on $\tan\beta$ for each $\lambda_d$. We take $\Lambda_D=1000$ TeV, and show the bounds for $M_{\rm mess}=10^7$ and $10^{12}$ GeV. At each point, $r_L$ is scanned to find a solution realizing successful EWSB. 
Below the solid lines, the stop becomes tachyonic with large $r_L$ and there is no solution to explain the EWSB.
It is found that for $\lambda_d=$0.9\,-1.0, $\tan\beta$ is required to be larger than about 9-\,10, while for more smaller $\tan\beta$, the constraint is much stronger than the Landau pole and tachyonic sneutrino constraint as shown in Fig.~\ref{fig:tachyonic_sneutrino}. Considering also this constraint, the upper bound on the Higgsino mass can be estimated roughly　
\begin{eqnarray}
|\mu| &\lesssim& (250, 400, 500)\,{\rm GeV}\,\, {\rm for}\,\, \Lambda_D = (1000, 2000, 4000)\,{\rm TeV}, 
 \label{mu_bound_0}
\end{eqnarray}
for $M_{\rm mess} \gtrsim 10^7$ GeV. As a result, it turns out that the Higgsino is always light. 

We emphasize that although the lower bound on $\tan\beta$ in Fig.~\ref{fig:neg_stop} is model dependent, it is generically true that small $\tan\beta$ with large $\lambda_d$ makes it difficult to be consistent with the correct EWSB. 
This is because the contribution to the Higgs potential $m_{H_d}^2/\tan^2\beta \propto \lambda_d^2/\tan^2\beta$ becomes large in such cases. 
Therefore, the light Higgsino is always favored together with the smaller fine-tuning.

\subsection{Mass spectra}

\begin{table*}[]
\begin{center}
\begin{tabular}{|c||c|c|c|c|}
\hline
Parameters & Point {\bf I} & Point {\bf II} & Point {\bf III} &  Point {\bf IV} \\
\hline
$\Lambda_{D}$ (GeV) & $8\cdot 10^5$  & $1.5\cdot 10^6$  & $3\cdot 10^6$ & $10^6$ \\
$M_{\rm mess}$ (GeV)& $5\cdot 10^6$ & $10^7$  & $10^7$ & $5\cdot 10^6$ \\
$\lambda_d$ & 0.8 & 0.7 & 0.39 & 0.9 \\
$\tan\beta$ & 10 & 10 & 6  & 13\\
\hline
$\mu$ (GeV) & -147.6 & -190.3  & -214.5 & -146.7\\
$r_L$ & $1.864$ & $1.664$  & $1.740$ & $1.517$\\
\hline
Particles & Mass (TeV) & Mass (TeV) & Mass (TeV) &  Mass (TeV) \\
\hline
$\tilde{t}_{1,2}$ & 5.5, 8.3 & 10, 15  & 21, 29 & 7.1, 9.7\\
$\tilde{g}$ & 5.4 & 9.6  & 18  & 6.6 \\
$m_{\tilde L}, m_{\tilde E}$ & 4.4, 4.4 & 7.6, 6.9  &17, 10 & 4.3, 4.8\\
$\tilde{\chi_2^\pm}$ & 4.0 & 6.6  & 14 & 4.0\\
$\tilde{\chi_3^0}$ & 1.7 & 3.0  & 6.4  & 1.9\\
$A$ & 37 & 55 & 65 & 43\\
\hline
$(h_{0})_{\rm FH} ({\rm GeV})$ & 125.4 &  128.4 & 133.1 & 127.6\\
$(h_{0})_{\rm SHD}  ({\rm GeV})$ & 122.7 &  125.2 & 125.9& 124.1\\
\hline
\end{tabular}
\caption{\small Mass spectra in sample points. Here, $\lambda_u$ and $r_L$ are determined by the EWSB conditions, and $(h_0)_{\rm FH}$ and $(h_0)_{\rm SHD}$ are the Higgs mass computed by {\tt FeynHiggs} and {\tt SUSYHD}, respectively.
}
\label{tab:sample}
\end{center}
\end{table*}

Finally, we present sample mass spectra in Table~1 ({\bf I}-{\bf IV}).
The input parameters are $M_{\rm mess}$, $\lambda_d$ and $\tan\beta$, and $r_L$ and $\lambda_u$ are fixed by the EWSB conditions. The MSSM mass spectra is calculated using {\tt SOFTSUSY}, 
and the Higgs boson mass calculated using {\tt FeynHiggs\,2.11.2}~\cite{feynhiggs} and {\tt SUSYHD}, denoted by $(h_0)_{\rm FH}$ and $(h_0)_{\rm SHD}$, respectively. For the point {\bf IV}, $(h_0)_{\rm SHD}$ is consistent with the observed Higgs boson mass including theoretical uncertainty about 1\,GeV.
Although the SUSY particles other than the Higgsino are heavy and beyond the reach of the LHC, 
the Higgsino is always light as $\mathcal{O}(100)$\,GeV. At all sample points, the lightest supersymmetric particle (LSP) and next-to-LSP are the gravitino and Higgsino, respectively.\footnote{
For exceptional cases, see e.g.\,~\cite{except1, except2}.
}

\section{Conclusion}

We have shown that a simple and naive way of generating the $\mu$ ($B_\mu$) term can work in the 10 TeV or heavier squark case with the small breaking of the GUT relation among messenger $B$-terms. 
The required doublet/triplet splitting, $\Lambda_L/\Lambda_D > 1$, may be naturally accommodated in more general gauge mediation models even if GUT relations among parameters in the messenger sector are satisfied~\cite{Meade:2008wd,Buican:2008ws}.

In this paper, we also investigate the upper bound on the Higgsino mass focusing on the $\mathcal{O}(10)$ TeV squark case. Such a squark mass region is favored by the naturalness of achieving the correct EWSB 
rather than a region of much heavier squarks.
The ratio between the two couplings of messenger-Higgs interactions, which generate the $\mu$ ($B_\mu$) term, is fixed by $\tan\beta$, and the size of the coupling is bounded from above by the conditions of not just keeping the perturbativity up to the GUT scale but also avoiding tachyonic sneutrinos. Then, the Higgsino mass is bounded from above for a fixed SUSY mass scale. (Here, $\tan\beta$ is determined by the observed Higgs boson mass.) 
Furthermore, in cases of the larger squark masses, 
the stronger upper bound on the messenger-Higgs coupling is imposed in order to satisfy the EWSB conditions. 
Consequently, it turns out that the light Higgsino, $|\mu| < 500$ GeV, is a promising signal of gauge mediation. Although the signal also depends on the Higgsino lifetime which can be taken broad, 
it can be accessible by the LHC~\cite{Meade:2010ji,LHC_higgsino_1,LHC_higgsino_2} and ILC~\cite{Berggren:2013vfa}.

\section{Acknowledgments}\label{sec:ackno}
We thank Yuichiro Nakai for useful discussions. 
This work is supported by the German Research Foundation through TRR33 ``The Dark Universe" (MA). 
The research leading to these results has received funding from the European Research Council under the European Unions Seventh Framework Programme (FP/2007-2013) / ERC Grant Agreement n. 279972 ``NPFlavour'' (NY).

\appendix

\section{Soft SUSY breaking masses with splitting messenger $B$-terms} \label{sec:mgmsb}

The relevant superpotential is given by
\begin{eqnarray}
W = ( k_L Z  + M_L ) \Psi_{L} {\bar \Psi}_{L}  +  (k_D Z  + M_D ) \Psi_{D} {\bar \Psi}_{D},
\end{eqnarray}
where $\Psi_L$ and $\Psi_D$ are $SU(2)_L$ doublet and $SU(3)_c$ triplet, respectively, and $U(1)_Y$ charges of $\Psi_L$ and $\Psi_D$ are taken as (1/2) and (-1/3). 

Then, gaugino masses are given by
\begin{eqnarray}
M_{\tilde b} \simeq \frac{g_1^2}{16\pi^2} (\frac{3}{5} \Lambda_L + \frac{2}{5} \Lambda_D ) , \ \ 
M_{\tilde w} \simeq \frac{g_2^2}{16\pi^2}   \Lambda_L, \ \ 
M_{\tilde g} \simeq \frac{g_3^2}{16\pi^2}   \Lambda_D,
\end{eqnarray}
where $\Lambda_L = k_L \left<F_Z\right>/M_L$ and $\Lambda_D = k_D  \left<F_Z\right>/M_D$.
The SM gauge couplings of $SU(3)_c$, $SU(2)_L$ and $U(1)_Y$ are denoted by $g_3$, $g_2$ and $g_1$.
Here, $k_{L,D} \left<Z\right> \ll M_{L,D}$ is assumed. Scalar masses are 
\begin{eqnarray}
m_{\tilde{Q}}^2 &\simeq& \frac{2}{(16\pi^2)^2}
\left[
\frac43 g_3^4 \Lambda_D^2
+ \frac34 g_2^4 \Lambda_L^2
+ \frac35 g_1^4 (\tilde \Lambda_1^2) \frac{1}{6^2}
\right],
\nonumber \\
m_{\tilde{U}}^2 &\simeq& \frac{2}{(16\pi^2)^2}
\left[
\frac43 g_3^4 \Lambda_D^2 + \frac35 g_1^4 (\tilde \Lambda_1^2)
\left(\frac23\right)^2
\right],
\nonumber \\
m_{\tilde{D}}^2 &\simeq& \frac{2}{(16\pi^2)^2}
\left[\frac43 g_3^4 \Lambda_D^2 + \frac35 g_1^4 (\tilde \Lambda_1^2) \frac{1}{3^2}
\right],
\nonumber \\
m_{\tilde{L}}^2 &\simeq& \frac{2}{(16\pi^2)^2}
\left[ \frac34 g_2^4 \Lambda_L^2+ \frac35 g_1^4 (\tilde \Lambda_1^2)
\frac{1}{2^2}
\right],
\nonumber \\
m_{\tilde{E}}^2 &\simeq& \frac{2}{(16\pi^2)^2}
\left[ \frac35 g_1^4 (\tilde \Lambda_1^2) \right],
\nonumber \\
m_{H_u}^2 &=& m_{H_d}^2 = m_{\tilde{L}}^2,
\end{eqnarray}
with $\tilde \Lambda_1^2 \equiv [(3/5) \Lambda_L^2  + (2/5) \Lambda_D^2] $.

\section{Renormalization group equations}
We list the RGEs for $\lambda_d$ and $\lambda_u$: 
\begin{eqnarray}
\beta_{\lambda_d} &=& \frac{\lambda_d}{16\pi^2} \left(4 \lambda_d^2 +3 Y_b^2 + Y_{\tau}^2 -3 g_2^2 -\frac{3}{5}g_1^2 \right) \,, \nonumber \\
\beta_{\lambda_u} &=& \frac{\lambda_u}{16\pi^2} \left(4 \lambda_u^2 +3 Y_t^2 -3 g_2^2 -\frac{3}{5}g_1^2 \right) \, ,
\end{eqnarray}
where we have neglected contributions from $k_D$, $k_L$ and Yukawa couplings of first and second generations.
Also, the beta-functions for MSSM Yukawa couplings are modified. The changes of beta-functions are 
\begin{eqnarray}
\delta \beta_{Y_t} &=& \frac{Y_t}{16\pi^2}  \, \lambda_u^2 \,,\nonumber \\
\delta \beta_{Y_b} &=& \frac{Y_b}{16\pi^2} \, \lambda_d^2 \,,\nonumber \\
\delta \beta_{Y_{\tau}} &=& \frac{Y_{\tau}}{16\pi^2} \, \lambda_d^2 \,.
\end{eqnarray}
Above the messenger scale, the beta-functions of gauge couplings have additional contributions:
\begin{eqnarray}
\delta \beta_{g_i} = \frac{g_i^3}{16\pi^2} N_{\rm mess},
\end{eqnarray}
where $N_{\rm mess}$ is a number of messenger superfields.

%
%

\end{document}